\documentclass[english]{maciarticle}

\usepackage{amsmath, amsbsy, amscd, amssymb, graphicx, epsfig, color, bbm, wrapfig}
\usepackage{tikz}
\usepackage{booktabs}
\usepackage{amsmath}
\usepackage{graphicx, float}
\usepackage[margin={1.2cm,1.2cm}, bmargin={2cm}, tmargin={2.2cm}]{geometry}
\usepackage{comment}

\begin{document}

\title{Option Pricing Model with Transaction Costs}

\author[1]{F. Bellora}
\author[2]{G. Mazzei}
\author[3]{M. Maurette}
\affil[1]{Carnegie Mellon University, 5000 Forbes Ave, Pittsburgh, PA 15213, United States, fbellora@andrew.cmu.edu}
\affil[2]{Universit$\acute{e}$ Paris-Saclay, 3 Rue Joliot Curie, 91190 Gif-sur-Yvette, France,  gaston.mazzei@universite-paris-saclay.fr}
\affil[3]{CRISIL GR\&A, A S\&P Company, Libertador 174 10$^{th}$ floor, Vicente Lopez, Buenos Aires Argentina, manuel.maurette@crisil.com}

\maketitle

\vspace{-1cm}
\begin{abstract}
The author presents alternatives to the Black-Scholes european call option pricing model by incorporating different transaction cost structures in the replicating strategy.
In particular, an exponentially decreasing structure is proposed and developed.
\end{abstract}
\begin{keywords}
option, pricing, transaction costs
\end{keywords}

{\thispagestyle{empty}} 

\section{Introduction}

The Black-Scholes european call option pricing model rests on the existence of a replicating portfolio strategy with a deterministic evolution and the use of a non-arbitrage argument \cite{hull}.
The main problem arising with the introduction of transaction costs, is that the replicating strategy presented by Black-Scholes, requires continous delta-hedging of the replicating portfolio, which ultimately leads to unbounded transaction costs.
This problem has been tackled by Leland \cite{leland} by the use of a discrete-time replicating strategy presented by Boyle and Emanuel \cite{boyle}.

Consider the following transaction cost function $\Upsilon (S |\nu|) = S |\nu|k \left(S |\nu|\right)$, where $S$ is the value of the underlying asset, $\nu$ the amount of traded stocks (therefore $S|\nu|$ is the total value of the transaction), and $k(S|\nu|)$, a function which represents a proportion of the total value of the transaction.
We will work with $\Upsilon (S |\nu|)$ as a nonincreasing, non-negative function, and $S$ an underlying asset which follows a geometric brownian motion with constant volatility $\sigma$.
We will consider the existence of a risk-free interest rate $r$.







Following the argument presented by Leland \cite{leland}, if $C(S,t)$ denotes the value of the european call option for an underlying price $S$ at time $t$, by using a replicating portfolio $\Pi = C - \Delta S$ and selecting an appropriate discrete-time delta hedging strategy $\Delta = C_{S}$, we obtain the nonlinear evolution equation for a european call option with transaction costs.
Where $\phi \sim N(0,1)$.

\begin{equation}
\label{posta}
  \frac{\partial C}{\partial t} + \frac{1}{2} \sigma^{2} S^{2} \frac{\partial^{2} C}{\partial S^{2}} - \frac{ \sigma S^{2} }{\sqrt{\delta t}} \left| \frac{\partial^{2} C}{\partial S^{2}} \right| \mathbb{E} \left( \hspace{1mm} \bigl|  \phi \hspace{1mm} k(S,|\nu|) \bigr|  \hspace{1mm} \right) + r S \frac{\partial C}{\partial S} - r C = 0.
\end{equation}

\noindent For \ref{posta} not to be ill-posed the following condition must be satisfied:

\begin{equation}
    \frac{1}{2} - \frac{\textrm{sg}( C_{SS})}{\sigma \sqrt{ \delta t}} \mathbb{E} \left( \hspace{1mm} \bigl|  \phi \hspace{1mm} k(S,|\nu|) \bigr|  \hspace{1mm} \right) > 0
\end{equation}

In the following sections we study via numerical methods \cite{duffy} the results of incorporating transaction costs for different functions $k=k(S|\nu|)$, under the following Dirichlet boundary conditions.

\vspace{-0.5cm}
\begin{equation}
\label{contorno}
   C(0,t) = 0 \hspace{0.3cm} \forall t \in [0,T], \hspace{0.7cm} C(S,T) = \textrm{max}(0, S(T) - K), \hspace{0.7cm} C(S,t) \xrightarrow[{S \to \infty }]{} S \hspace{0.15cm} ; \hspace{0.15cm} 0 \leq t \leq T
\end{equation}

\noindent The model we consider is for cases in which the underlying asset price $S$ is around the strike price $K$. 

\vspace{-0.2cm}
This work is organized as follows: In Section \ref{escalonado} we study a decreasing piecewise constant function, which is most common in today's markets; in Section \ref{exp} we will study a function whose proportional costs are exponentially decreasing.
Section \ref{conclu} is a comparison between different cost structures.
Section \ref{apendice} contains the proof of a proposition regarding exponentially decreasing functions.

\section{Piecewise Constant Proportion Transaction Costs}
\label{escalonado}

In this section we study a cost structure in which the proportion $k$ on the total transaction is a piecewise constant decreasing function.
This is the most common cost structure offered by finance brokers around the world.
The following is an example of the cost structures used in online brokers \textit{Etrade} and \textit{TradeStation}.

\begin{table}[H]
\centering
\begin{tabular}{*3l}    
\toprule
Amount of Transactions & ETrade  Cost (USD) & TradeStation  Cost (USD)\\\midrule

  \hspace{0.2cm} $[0, 50)$ & $10$ & $9$  \\
  \hspace{0.2cm} $[50,200)$ & $10$ & $7$ \\
  \hspace{0.2cm} $[200, \infty)$ & $8$ & $5$ \\

 \bottomrule
\end{tabular}
\caption{\textit{Consulted on November 2016}}
\label{brokers}
\end{table}

A particular case, contained within this function is the one studied by Leland \cite{leland} in which he considers the transaction cost to be a constant proportion of the total traded value, i.e. $\Upsilon (S|\nu|)=kS|\nu|$.
In this case we consider the following cost function:

\begin{equation}
\label{escalon_cost}
  \Upsilon (S|\nu|)=\sum_{i=1}^n{} \mathbbm{1}_{\{[x_{i-1}, x_{i}]\}} k_{i} S|\nu| \hspace{3mm} ; \hspace{2mm} k_{1} \geq ... \geq k_{n} \hspace{2mm} ; \hspace{2mm} 0 = x_{0} < ... < x_{n}=\infty.
\end{equation}

Substituing in (\ref{posta}) we obtain the following system of non-linear partial differential equations, where the one used will depend on the value of $S|\nu|$ at each point in time.

\begin{equation}
  \label{Antidoto}
  \begin{cases}
  \vspace{3mm}
     \frac{\partial C}{\partial t} + \frac{1}{2} \sigma^{2} S^{2} \frac{\partial^{2} C}{\partial S^{2}} -  k_{1} \sigma S^{2} \sqrt{\frac{2}{\pi \delta t}} \left| \frac{\partial^{2} C}{\partial S^{2}} \right|  + r S \frac{\partial C}{\partial S} - r C = 0  \hspace{4mm} \text{if} \hspace{2mm} 0 <S|\nu|< x_1
  \\
  ...
  \\
  \frac{\partial C}{\partial t} + \frac{1}{2} \sigma^{2} S^{2} \frac{\partial^{2} C}{\partial S^{2}} -  k_{n} \sigma S^{2} \sqrt{\frac{2}{\pi \delta t}} \left| \frac{\partial^{2} C}{\partial S^{2}} \right|  + r S \frac{\partial C}{\partial S} - r C = 0  \hspace{4mm} \text{if} \hspace{2mm} x_{n-1} < S|\nu|
  \end{cases}
\end{equation}

\begin{figure}[h!]
\begin{center}
\includegraphics[width=17cm, keepaspectratio]{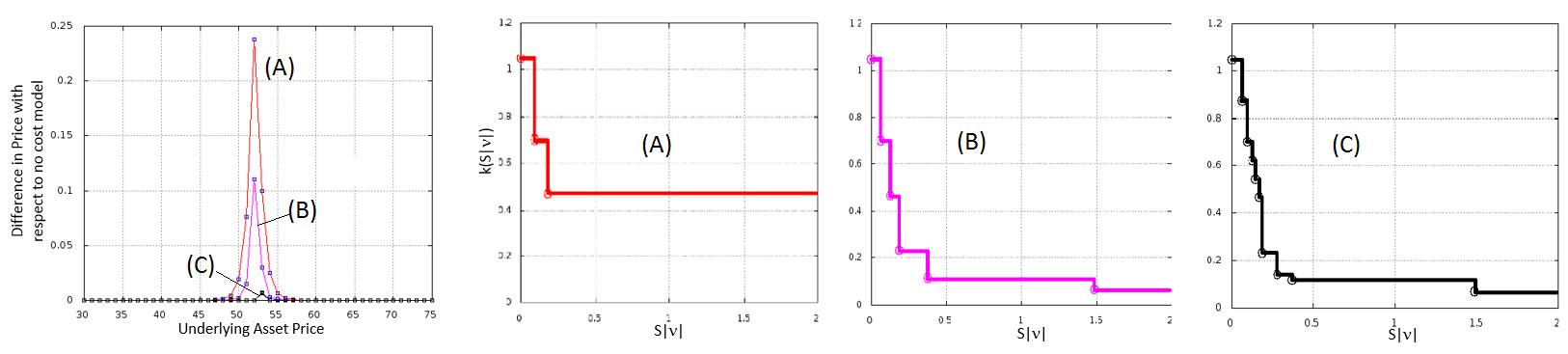}
\caption{The left-most graph shows the difference in price between the model without transaction costs and different piecewise constant models with transaction costs.  The other three plots (A),(B) and (C) are the  different piecewise constant proportions $k(S|\nu|)$ used.} \label{grafico_esc}
\end{center}
\end{figure}

The above was implemented using backward finite-difference method with (\ref{contorno}) as boundary conditions.
We observe in Figure \ref{grafico_esc} that the introduction of these type of transaction costs to the model, will yield different results when the option is \textit{at the money}.
However if the underlying asset price is such that $S<<K$ or $S>>K$ there is no noticeable difference between the two models.
This is because, the cost term in \ref{Antidoto} depends on $\Gamma = C_{SS}$ which tends to zero when the option isn't at the money (this will be shown later in Fig. \ref{grafico_exp}).
This means that when the option is extremely out of the money or in the money there will be very little rehedging.
It is possible, if the cost is large enough, for the option price to become negative, this will occur around the strike price $K$.

\vspace{-1cm}
In Figure \ref{grafico_esc}, we also observe that by partitioning the steps of the function $k(S|\nu|)$, the difference between the model without and with transaction costs is lower.
This result is expected given the condition $k_{1} \geq ... \geq k_{n}$ which states that as we add steps to the function, transaction costs will be lower.

\section{Exponentially Decreasing Transaction Costs}
\label{exp}

In this section we study a cost structure in which the proportion $k$ on the total transaction is exponentially decreasing.
We will use the following cost function:

\begin{equation}
\label{costoexp}
 \Upsilon (S|\nu |) =  S|\nu | e^{-(a+S|\nu |)}.
\end{equation}

The choice of using (\ref{costoexp}) relies upon the properties this function has.
In first place it is non-negative (as opposed to a linear proportion), its continuous (as opposed to a piecewise constant proportion), and it decreases when the total transaction is bigger.
As far as we know, this work is the first one this transaction cost structure is proposed and studied, we study the particular case of long vanilla calls.
In order to introduce this cost function in (\ref{posta}) we shall make use of Proposition \ref{esperanza}.

\vspace{-0.3cm}
\begin{proposition}
\label{esperanza}
(Proof in Appendix) \hspace{5mm} Let $\phi \sim N(0,1)$ and $\alpha \in \mathbb{R}$, then:

\begin{equation}
\label{esp}
  \mathbb{E} \left[ | \phi | e^{-\alpha |\phi|} \right] = \sqrt{\frac{2}{\pi}} - 2 \alpha \hspace{0.5mm}\Phi( - \alpha ) \hspace{0.5mm} \textrm{exp}{\left( \frac{\alpha^{2}}{2} \right)}.
\end{equation}
\end{proposition}

Using (\ref{esp}) we can substitute in (\ref{posta}) obtaining the following deterministic, non-linear PDE, to price european calls under a cost structure given by (\ref{costoexp}).

\begin{equation}
\label{eqexp}
\begin{aligned}
0 = C_{t} + \frac{1}{2} \sigma^{2} S^{2} C_{SS} + r S C_{S} - r C - 
  \sqrt{\frac{2}{\pi \delta t}} \sigma S^{2} e^{-a} \left| C_{SS} \right| +
  \\ 2 \left| C_{SS} \right|^{2} \sigma^{2} S^{4} e^{-a} \hspace{1mm} \textrm{exp} \left( \frac{1}{2} \left| C_{SS} \right|^{2} \sigma^{2} S^{4} \delta t \right)
  \Phi \left( - \left| C_{SS} \right| \sigma S^{2} \sqrt{\delta t} \right).  
\end{aligned}
\end{equation}

\begin{figure}[h!]
\begin{center}
\includegraphics[width=14cm, height=5cm]{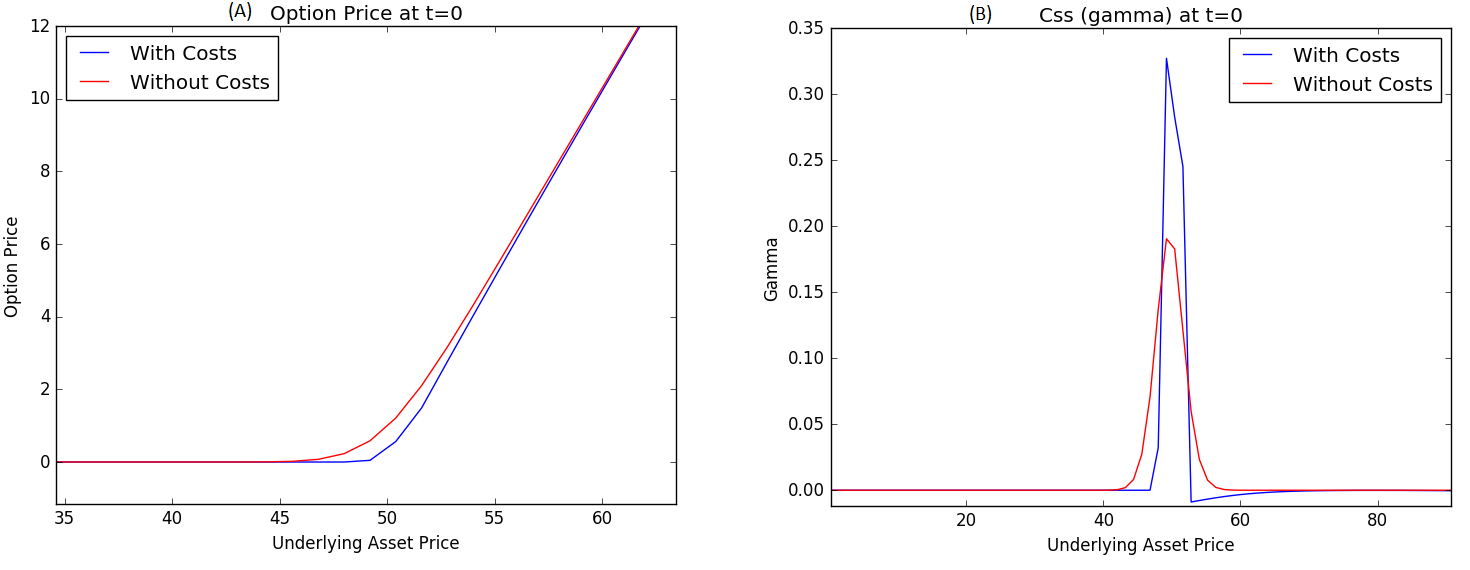}
\caption{\textbf{Left:} Option Price at $t=0$. \textbf{Right:} 
$\Gamma = C_{SS}$ at $t=0$, as a function of the underlying asset price $S$.} \label{grafico_exp}
\end{center}
\end{figure}

The above was implemented using backward finite-difference method with (\ref{contorno}) as boundary conditions.  
In Figure \ref{grafico_exp}, we observe that the introduction of transaction costs to the replicating strategy, has a large impact when the option is \textit{at the money}, making its price lower given that one will incur in certain costs to maintain the replication strategy.

\vspace{-1cm}
The reason for this difference between the two models (with/without costs) is that the transaction cost arises from the rehedging of a previously delta-hedged portfolio.
This means that the corrections to be made in the replicating strategy arise from $\Gamma = C_{SS}$.
In the right hand plot in Figure \ref{grafico_exp} we can observe that with the introduction of transaction costs, $\Gamma$ conserve its peak-like shape but it reaches a higher maximum value.
For this particular case (long vanilla calls) we observe that $\Gamma$ is a positive function.

\section{Comparison Between the Different Cost Structure Models}
\label{conclu}

%

%

Upon comparing different cost structures, of the form $\Upsilon (S|\nu|) = S|\nu| k(S|\nu|)$, we can observe that in general, the option price will be lower around the Strike Price.

\begin{wrapfigure}{l}{11cm}
    \includegraphics[width=10.5cm, height=8cm]{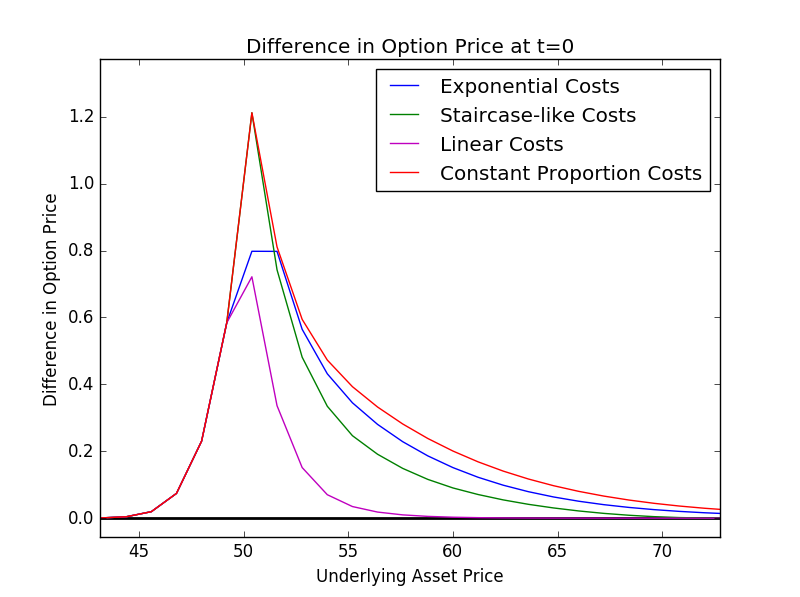}
    \vspace{-0.5cm}
    \caption{Difference in option price at $t=0$ between various transaction cost structures and Black-Scholes model without transaction costs.} \label{c_todos}
\end{wrapfigure}

In Figure \ref{c_todos} we compare different cost structures, where we include:
Staricase-like decreasing proportional costs, along with the particular case of constant proportion costs; linearly decreasing proportional costs (studied by Amster \cite{amster}) considering a function of the form $\Upsilon (S|\nu|) = (a - kS|\nu|) S|\nu|$ (Maurette \cite{maurette} did a numerical analysis on this cost structure) and finally exponentially decreasing transaction costs.
In order to compare the different cost structures, the parameters within the cost function have been adjusted so that $k(0)=0.05$.

\newpage

It can be seen that, around the Strike Price, the difference between the models which incorporate transaction costs and the model without transaction costs, is larger for the constant proportion costs, and the piecewise constant cost structure.
Constant proportion costs, produce the biggest difference between the models because its the most expensive cost structure given that there is no benefit upon trading larger amounts of stock.
In the, \textit{in the money} region, we observe that the difference between the models approaches zero, however drawing conclusions about which cost structure approaches zero faster, depends entirely upon the particular function used.
An example of this is that the exponential cost structure, appears to fall below the constant proportion but above the piecewise constant function, however as stated before, the constant proportion is just a particular case within the piecewise function.
As for the linearly decreasing cost structure, we can see that it exhibits the same behaviour as the other cost structures studied in this work.

These results invite to continue further investigations on transaction costs, proposing new structures and obtaining stronger results on the ones presented in this work.

\vspace{-0.3cm}
\section{Appendix}
\label{apendice}

\begin{proof} (Proposition \ref{esperanza}):
Let $Z \sim \mathbf{N}(\mu,\sigma)$, $a \in{} \mathbb{R}$  we want to find $\mathbb{E} \left[ | z | e^{-a |z|}  \right]$.
In order to obtain an expression for this, we will use the linearity property of the expected vale operator.
\begin{equation}
\label{daE}
    \mathbb{E} \left[ | z | e^{-a |z|} \right] = \mathbb{E} \left[ - \frac{{\partial }}{{\partial a}} e^{-a |z|} \right] = - \frac{{\partial }}{{\partial a}} \mathbb{E} \left[ e^{-a |z|} \right].
\end{equation}

We will then proceed to calculate $\mathbb{E} \left[ e^{-a |z|} \right]$ by definition.
\begin{equation}
    \mathbb{E} \left[ e^{-a |z|} \right] = \int_{-\infty}^{\infty} e^{-a |z|} \frac{1}{\sqrt{2 \pi} \sigma} e^{-\frac{(z-\mu)^2}{2 \sigma^2}} dz.
\end{equation}


Using a simple completing squares method we obtain:


\begin{equation*}
    \mathbb{E} \left[ e^{-a |z|} \right] = 
           \int_{-\infty}^{0} e^{\frac{a(2\mu + a \sigma^2)}{2}} \frac{1}{\sqrt{2 \pi} \sigma} e^{-\frac{1}{2} \left( \frac{z-(\mu+a\sigma^2)}{\sigma} \right) ^{2}} dz
           \int_{0}^{\infty} e^{- \frac{a(2\mu - a \sigma^2)}{2}} \frac{1}{\sqrt{2 \pi} \sigma} e^{ -\frac{1}{2} \left( \frac{z-(\mu-a\sigma^2)}{\sigma} \right) ^{2}} dz,
\end{equation*}

substituing $y_{1} = \frac{z-(\mu+a\sigma^2)}{\sigma}$ \hspace{0.5cm};\hspace{0.5cm}  $y_{2} = \frac{z-(\mu-a\sigma^2)}{\sigma}$ we obtain:

\begin{equation*}
    \mathbb{E} \left[ e^{-a |z|} \right] = 
     e^{\mu + \frac{ a^{2} \sigma^{2}}{2}}  \int_{-\infty}^{- \frac{\mu + a \sigma^{2}}{\sigma}} \frac{1}{\sqrt{2 \pi}} e^{-\frac{y_{1}^{2}}{2}} dy_{1}  \hspace{2mm} + \hspace{2mm} 
     e^{-\mu + \frac{ a^{2} \sigma^{2}}{2}}  \int_{- \frac{\mu - a \sigma^{2}}{\sigma}}^{\infty} \frac{1}{\sqrt{2 \pi}} e^{-\frac{y_{2}^{2}}{2}} dy_{2}.
\end{equation*}

Let $\Phi (z) = \mathbb{P} [Z \leq z] = \int_{-\infty}^{z}   \frac{1}{\sqrt{2 \pi}} e^{-\frac{y^{2}}{2}} dy$ \hspace{0.3cm} then,

\begin{equation*}
   \mathbb{E} \left[ e^{-a |z|} \right] =  e^{\frac{ a^{2} \sigma^{2}}{2}} \left[ e^{\mu} \hspace{1.3mm} \Phi \left(- \frac{\mu + a \sigma^{2}}{\sigma} \right)
+ e^{-\mu} \hspace{1.3mm} \Phi \left(  \frac{ \mu - a \sigma^{2}}{\sigma} \right) \right].
\end{equation*}

Once obtained the expression for $\mathbb{E} \left[ e^{-a |z|} \right]$ we use the property mentioned in (\ref{daE}), from where we arrive at the following expression for $\mathbb{E} \left[ | z | e^{-a |z|} \right]$ where $Z \sim \mathbf{N}(\mu,\sigma)$.




\begin{equation}
\mathbb{E} \left[ | z | e^{-a |z|} \right] =   - a \sigma^{2} e^{\frac{ a^{2} \sigma^{2}}{2}} \left[ e^{\mu} \hspace{1.3mm} \Phi \left(- \frac{\mu + a \sigma^{2}}{\sigma} \right) + e^{-\mu} \hspace{1.3mm} \Phi \left(  \frac{ \mu - a \sigma^{2}}{\sigma} \right) \right] + 
\sigma \sqrt{\frac{2}{\pi}} e^{ - \frac{\mu^{2}}{2 \sigma^{2}}} \hspace{1.3mm} \textrm{cosh} [ \mu (1 - a) ].
\end{equation}

In Section \ref{exp} we will use this result for the particular case in which $\phi \sim \mathbf{N}(0,1)$; $\alpha \in \mathbb{R}$. 
For this case we obtain:

\begin{equation}
  \mathbb{E} \left[ | \phi | e^{-\alpha |\phi|} \right] = \sqrt{\frac{2}{\pi}} - 2 \alpha \hspace{0.5mm}\Phi( - \alpha ) \hspace{0.5mm} \textrm{exp}{\left( \frac{\alpha^{2}}{2} \right)}.  
\end{equation}
\end{proof}

\vspace{-1cm}
\section*{Acknowledgements}

The author thanks Andr\'es Mogni and Manuel Maurette for their insight on the interpretation of results.

\vspace{-0.3cm}

\end{document}